\documentclass[prb,twocolumn,a4paper,superscriptaddress,floatfix,showpacs]{revtex4}
\usepackage{float}
\usepackage{amsmath}
\usepackage{amssymb}
\usepackage{graphicx}
\usepackage{color}
\usepackage{bm}
\usepackage{url}
\usepackage{epstopdf}
\begin{document}

\title{Localized electron states near the armchair edge of graphene}

\author{P.\,A.~Maksimov}
\thanks{Corresponding author. E-mail: maksimov.pavel@yahoo.com}
\affiliation{Moscow Institute of Physics and Technology, Institutskiy per. 9, Dolgoprudnyi, Moscow Region, 141700 Russia}
\affiliation{Institute for Theoretical and Applied Electrodynamics, Russian Academy of Sciences, 125412 Moscow, Russia}

\author{A.\,V.~Rozhkov}
\affiliation{Moscow Institute of Physics and Technology, Institutskiy per. 9, Dolgoprudnyi, Moscow Region, 141700 Russia}
\affiliation{Institute for Theoretical and Applied Electrodynamics, Russian Academy of Sciences, 125412 Moscow, Russia}

\author{A.\,O.~Sboychakov}
\affiliation{Institute for Theoretical and Applied Electrodynamics, Russian Academy of Sciences, 125412 Moscow, Russia}

\begin{abstract}
It is known that zigzag graphene edge is able to support edge states: There
is a non-dispersive single-electron band localized near the zigzag edge.
However, it is generally believed that no edge states exist near unmodified
armchair edge, while they do appear if the edge is subjected to suitable
modifications (e.g. chemical functionalization). We explicitly present two
types of the edge modification which support the localized states. Unlike
zigzag edge states, which have zero energy and show no dispersion,
properties of the armchair localized states depend sensitively on the type
of edge modification. Under suitable conditions they demonstrate pronounced
dispersion. While the zigzag edge state wave function decays monotonously
when we move away from the edge, the armchair edge state wave function shows
non-monotonous decay. Such states may be observed in scanning tunneling
spectroscopy experimentally.
\end{abstract}


%73.22.Pr 	Electronic structure of graphene
%73.20.At	Surface states, band structure, electron density of states

\date{\today}

\maketitle
\section{Introduction}
Investigation of graphene sheet edges is an active area of the condensed
matter research
\cite{meso_review,review_mag_yazaev,neto_etal}.
The importance of these studies stems from the fact that the electronic
properties of mesoscopic objects depend sensitively on properties of their
edges. For example, in the case of graphene nanoribbons, modifications to
the edge chemistry
\cite{cervantes_func_mag,pisani_monohydrogen,our_nanoribbon_paper_2009,
gunlycke_edge_funct}
or introduction of the edge disorder
\cite{gunlycke_edge_disorder,areshkin_transport,
evaldsson_edge_disorder_nribb,waka_klein}
affect nanoribbon's transport or magnetic properties.

For a finite graphene sheet, there are two highly symmetric types of
termination: armchair and zigzag. It has been known since mid-90's
\cite{nakada-fujita_nribb_edge_st,fujita_nanoribbon_hubbard,
klein_edge}
that the zigzag edge binds electrons: there is a non-dispersive
single-electron band localized near the edge. Experimental data,
corroborating this theoretical idea, are also available
\cite{niimi_ldos,Tao2011}.

As for the armchair termination, edge states can be found in the presence
of magnetic field~\cite{wkb_magst,current_magst,ukr_magst,dehydr_magst}.
%{\color{red}
%after structural reconstruction~\cite{changwon} or with the external
%potential applied to atoms at the edge~\cite{klos}.
It is generally believed that without magnetic field pristine edge cannot
support a localized-state band~\cite{no_edge_state_armchair2012}.
The absence of such a band was demonstrated
experimentally~\cite{enoki_experiment}.
%}
However, this statement, as will be shown below, needs certain
qualifications. Indeed, the common model for $\pi$-electrons near the
armchair termination does not allow for the edge
states~\cite{no_edge_state_armchair2012}.
This property is a consequence of the two implicit assumptions built into
the latter model. Namely, it is postulated that {\it (i)} the hopping
integrals $t$ between the nearest-neighbor carbon atoms are the same both in
the bulk of the sample and near the edge, and {\it (ii)} no non-carbon
atoms or functional groups are attached to the unsaturated chemical bonds
at the edge.

However, these conditions are likely to oversimplify the reality. Using
density functional theory calculations it has been demonstrated that the
hopping amplitude between the carbon atoms at the edge differs from the
hopping amplitude in the bulk~\cite{son_gap}.
%Also, numerical
%calcualtions~\cite{li_t0} predict localized states in nanoribbon where this
%hopping amplitude at the edge is less than in the bulk. We expand these
%results to a broader range of parameters and present analytical approach to
%this problem.
Regarding condition {\it (ii)}, since atoms at the edge have unsaturated
bonds, one can imagine a situation in which chemical radicals (functional
groups or atoms) are attached to saturate these bonds. If $\pi$-electron
from graphene can hybridize with orbitals on attached radicals, these
non-carbon orbitals must be included into the model Hamiltonian. Such
orbitals appear in the Hamiltonian as extra sites at the edge, which are
connected by electron hopping to nearest-neighbor carbon atoms.  Of course,
the corresponding hopping amplitude differs from $t$, and the on-site
potential for these extra sites does not necessary equal to the Fermi
energy of graphene.

The violation of either {\it (i)} or {\it (ii)} has important consequences
for the physics of the armchair edge. Loosely speaking, when either of
these assumptions does not hold true, boundary conditions for an electron
wave function change, which might result in stabilization of the localized
solutions of the corresponding Schr\"odinger equation.
Indeed, the emergence of the localized states was reported in numerical
studies of the modified armchair
edge~\cite{li_t0,changwon,klos}.

In this paper we develop an analytic method for systematic investigation
of the localized bands at the armchair termination. Two types of edge
modifications will be studied. For both types we will determine the
conditions of the edge state stabilization. We will compare our findings
with the previous numerical results.
%Refs.~\onlinecite{li_t0,changwon,klos}.
%Another type of the edge band, which was not mentioned in the latter
%references, will be also described.

The paper is organized as follows. In Sec.~\ref{solution} we construct localized state wave function. In Sec.~\ref{section_dt} the edge with modified hopping integral is studied, and in Sec.~\ref{section_radical} we study the functionalized armchair edge. The results are discussed in Sec.~\ref{discussion}. Additional technical details are given in two Appendices.

\section{The solution of the Schr\"odinger equation in the form of decaying
wave}
\label{solution}

\begin{figure}
\includegraphics[width=0.9\columnwidth]{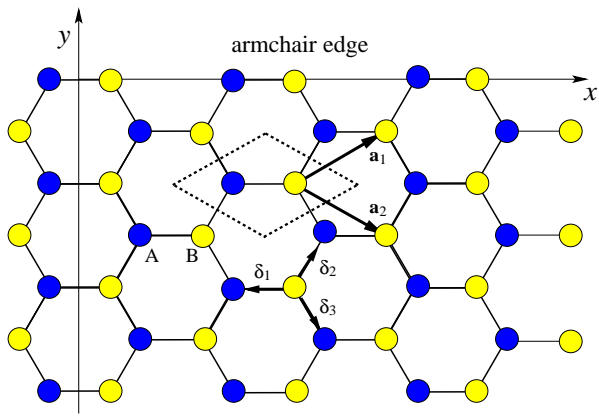}
\caption{(Color online) Honeycomb graphene lattice and armchair edge. Sublattice A is blue (dark-grey), sublattice B is yellow (light-grey). Primitive lattice vectors $\bm{a}_1,\bm{a}_2$ have coordinates $\frac{a_0}{2}(3,\pm \sqrt{3})$. Nearest-neighbor vectors: $\bm{\delta}_1=a_0(-1,0), \bm{\delta}_{2,3}=\frac{a_0}{2}(1,\pm \sqrt{3})$. Armchair edge is located at $y=0$ line. Sample is located in $y \leq 0$ half-plane.}
\label{graphene_sheet}
\end{figure}

In this paper we assume that electrons in the bulk of the graphene sheet are described by the tight-binding Hamiltonian with nearest-neighbor hopping:
\begin{equation}\label{H}
H=-t\sum_{\langle i,j \rangle,\sigma} {a}_{i,\sigma}^{\dag} {b}_{j,\sigma}+H.c.,
\end{equation}
where $a_{i,\sigma}$ ($b_{i,\sigma}^{\dag}$) are annihilation (creation) operators of electrons on the site $i$ of sublattice $A$ ($B$) with spin projection $\sigma$, and sum is taken over pairs of nearest-neighbor sites.
The coordinate axes are shown in
Fig.~\ref{graphene_sheet}.
The lattice vectors are equal to:
$\bm{a}_{1,2}=\frac{a_0}{2}(3,\pm \sqrt{3})$,
while the vectors connecting the nearest-neighbor atoms are given by the
following expressions:
$\bm{\delta}_1=a_0(-1,0),
\bm{\delta}_{2,3}=\frac{a_0}{2}(1,\pm \sqrt{3})$.
Here $a_0$ is the carbon-carbon bond length. Armchair edge is located at
$y=0$ line, and the sample is located in the $y \leq 0$ half-plane. In this
case, Schr\"odinger equation corresponding to
Hamiltonian~\eqref{H}
is
\begin{equation}
\begin{array}{r}
\varepsilon a(x,y) = -t b(x,y) -t b(x-\frac{3a_0}{2},y+\frac{\sqrt 3 a_0}{2}) \\
-t b(x-\frac{3a_0}{2},y-\frac{\sqrt 3 a_0}{2}),\\
\\
\varepsilon b(x,y) = -t a(x,y) -t a(x+\frac{3a_0}{2},y+\frac{\sqrt 3 a_0}{2}) \\
-t a(x+\frac{3a_0}{2},y-\frac{\sqrt 3 a_0}{2}),
\end{array}
\label{schrodmain}
\end{equation}
where $a(x,y)$, $b(x,y)$ are the components of the spinor
\begin{equation}
\Psi(x,y)=\left(\begin{array}{c}a(x,y)\\b(x,y)\end{array}\right).
\end{equation}
In this section we demonstrate that Schr\"odinger equation \eqref{schrodmain} admits a solution which propagates along $x$-axis as a plane wave and decays along $y$-axis exponentially:
\begin{equation}
\Psi_{\bm{k},\varkappa}(x,y)=\left(\begin{array}{c}\alpha_{\bm{k},\varkappa}\\\beta_{\bm{k},\varkappa}\end{array}\right)e^{ik_x x+ik_y y +\varkappa y}.
\label{wfac}
\end{equation}
Obviously, if exists, such solution cannot be a valid wave function in the bulk, since it is not normalizable. However, it can describe an edge state.

Eigenfunction $\Psi$ from Eq.~\eqref{wfac} is characterized by quasimomenta
$k_{x,y}$, inverse localization length $\varkappa$, and eigenenergy
$\varepsilon$. Not all of these four parameters are independent: below we
will show that, for $\Psi$ to serve as a Schr\"odinger equation solution, two conditions on the parameters have to be imposed. This reduces the number of independent parameters to two. Due to symmetries of our Hamiltonian, $\varepsilon$ and $k_x$ are conserved quantities, while $k_y$ and $\varkappa$ are not. Thus, it is convenient to label the wave function by $k_x$ and $\varepsilon$, and treat $\varkappa$ and $k_y$ as functions of $k_x$ and $\varepsilon$.

For wave function given by Eq.~\eqref{wfac} Schr\"odinger equation~\eqref{schrodmain} reads:
\begin{equation}
\varepsilon \left( \begin{array}{c}
\alpha\\
\beta
\end{array} \right) = \left( \begin{array}{ccc}0 & p\\ q & 0\end{array}\right) \left( \begin{array}{c}
\alpha\\
\beta
\end{array} \right),
\label{schac}
\end{equation}
\begin{equation}
\begin{array}{c}
p=-t \left[ 1+e^{-i\bar{k}_x}\left(e^{i{\bar{k}_y}+\bar{\varkappa}}+e^{-i{\bar{k}_y}-\bar{\varkappa}}\right)\right], \\
q=-t\left[ 1+e^{i\bar{k}_x}\left(e^{i{\bar{k}_y}+\bar{\varkappa}}+e^{-i{\bar{k}_y}-\bar{\varkappa}}\right)\right],\\
\end{array}
\label{pq}
\end{equation}
where, to simplify notation, we define dimensionless quantities $\bar{\varkappa},~\bar{k}_{x,y}$:
\begin{equation}
\bar{k}_x =\frac{3}{2} k_x a_0, ~\bar{k}_y= \frac{\sqrt{3}}{2} {k}_y a_0, ~\bar{\varkappa} =\frac{\sqrt{3}}{2}\varkappa a_0.
\end{equation}
Eigenvalues $\varepsilon$ of the matrix in Eq.~\eqref{schac} are given by:
\begin{equation}
\varepsilon^2=p q.
\end{equation}
For arbitrary $\bar{\varkappa},~\bar{k}_{x,y}$ this equation defines complex $\varepsilon^2$:
\begin{align}
\varepsilon^2 / t^2 &= 3+2\cos 2 {\bar{k}_y} \cosh 2 \bar{\varkappa} \nonumber \\
&+4\cos \bar{k}_x \cos {\bar{k}_y} \cosh \bar{\varkappa} {} \nonumber \\
& +  i 4\sin \bar{k}_x \sinh \bar{\varkappa} \left( \cos \bar{k}_x +2 \cos {\bar{k}_y} \cosh \bar{\varkappa} \right).\label{epsilon}
\end{align}
For $\varepsilon$ to be real (and to have the meaning of eigenenergy), we must impose two conditions:
\begin{align}
{\rm Im}\ \varepsilon^2 &= 4 t^2 \sin {\bar{k}_y} \sinh \bar{\varkappa} (\cos \bar{k}_x \nonumber \\ &+ 2\cos {\bar{k}_y} \cosh \bar{\varkappa})=0,
\label{ime}
\end{align}
\begin{align}
{\rm Re}\ \varepsilon^2 =t^2 (3 &+ 2\cos 2 {\bar{k}_y} \cosh 2 \bar{\varkappa} \nonumber \\  &+ 4\cos \bar{k}_x \cos {\bar{k}_y} \cosh \bar{\varkappa})>0.
\label{ree}
\end{align}
In addition, since we are interested in finding localized states, we add the third condition:
\begin{equation}
\bar{\varkappa}>0.
\label{kappapositive}
\end{equation}
Let us look closer at Eq.~\eqref{ime}. Since $\bar{\varkappa} \neq 0$, Eq.~\eqref{ime} is satisfied when either
\begin{equation}
\cos \bar{k}_x + 2\cos {\bar{k}_y} \cosh \bar{\varkappa}=0,
\label{ime2}
\end{equation}
or
\begin{equation}
\sin \bar{k}_y=0
\label{ime1}
\end{equation}
is valid.

Equations \eqref{ime2} and \eqref{ime1} cannot be simultaneously satisfied. If the wave function parameters satisfy Eq.~\eqref{ime2} we will refer to such a wave function as 'type A'. When Eq.~\eqref{ime1} is satisfied we refer to the wave function as 'type B'. Since properties of A and B types are quite dissimilar, we will treat each type separately.

\subsection{Type A solution: Eq.~\eqref{ime2} is satisfied}

When Eq.~\eqref{ime2} is valid, the eigenenergy is given by the following expression:
\begin{equation}\label{ree2}
\varepsilon^2={\rm Re}\ \varepsilon^2 =t^2 \left(5-4\cosh^2 \bar{\varkappa} -\frac{\cos^2 \bar{k}_x}{\cosh^2 \bar{\varkappa}} \right).
\end{equation}
Simple calculations show that for type A wave function $|\varepsilon|$ is bounded from above:
\begin{equation}
|\varepsilon|<t.
\label{enAbound}
\end{equation}
Equations \eqref{ime2} and \eqref{ree2} define $|k_y|$ and $\varkappa$ as implicit functions of two conserved quantities, $\varepsilon$ and $k_x$. It is easy to demonstrate that for given $\varepsilon$ and $k_x$ only one value of $\varkappa$ satisfying Eq.~\eqref{kappapositive} is possible. To prove this, imagine that there are $\varkappa$ and $\varkappa'$ both satisfying Eq.~\eqref{ree2}. Then:
\begin{equation}
4\cosh^2 \bar{\varkappa} +\frac{\cos^2 \bar{k}_x}{\cosh^2 \bar{\varkappa}}=4\cosh^2 \bar{\varkappa}' +\frac{\cos^2 \bar{k}_x}{\cosh^2 \bar{\varkappa}'}.
\end{equation}
One can see that $\bar{\varkappa}=\bar{\varkappa}'$. However, for $k_y$, only its absolute value, but not its sign, is uniquely specified [see Eq.~\eqref{ime2}].
Therefore, a superposition with arbitrary coefficients $C_{1,2}$
\begin{equation}
\Psi=C_1 \left(\begin{array}{c}\alpha_1\\\beta_1\end{array}\right)e^{ik_x x+ik_y y +\varkappa y}+C_2 \left(\begin{array}{c}\alpha_2\\\beta_2\end{array}\right)e^{ik_x x-ik_y y +\varkappa y}
\label{wf2}
\end{equation}
is the most general type A wave function, with given $k_x$ and $\varepsilon$.

It is interesting to note that, unlike zigzag edge state, which decays monotonously as we move away from the edge, wave function \eqref{wf2} demonstrates non-monotonous (oscillating) decay.

\subsection{Type B solution: Eq.~\eqref{ime1} is satisfied}

For type B wave function Eq.~\eqref{ime1} is valid. It has two solutions inside the Brillouin zone: $\bar{k}_{y_1}=0$ and $\bar{k}_{y_2}=\pi$. Eigenenergies for these wave functions:
\begin{equation}\label{energyB}
\varepsilon_{1,2}^2=t^2 \left[ 3 + 2 \cosh 2 \bar{\varkappa}_{1,2} \pm 4\cos \bar{k}_x \cosh \bar{\varkappa}_{1,2} \right]
\end{equation}
are bounded from below:
\begin{equation}
|\varepsilon_{1,2}|>t\,.
\label{enBbound}
\end{equation}
If we apply the condition for equal energies of these solutions, $\varepsilon_1=\varepsilon_2=\varepsilon$, we obtain the following relation between $\bar{\varkappa}_1$ and $\bar{\varkappa}_2$:
\begin{equation}\label{condkapB}
\cos \bar{k}_x = \cosh \bar{\varkappa}_2-\cosh \bar{\varkappa}_1\,.
\end{equation}
In other words, two wave functions with identical $\varepsilon$ and $k_x$ may have unequal $\varkappa$'s. This means that the superposition with arbitrary coefficients $C_{1,2}$
\begin{equation}
\Psi =C_1 \left(\begin{array}{c}\alpha_1\\\beta_1 \end{array}\right)e^{i k_x x +\varkappa_1 y} + C_2 \left(\begin{array}{c}\alpha_2\\\beta_2 \end{array}\right) e^{ik_x x +i\frac{2\pi}{a_0\sqrt{3}}y+ \varkappa_2 y}
\label{wf1}
\end{equation}
is the most general type B solution of the bulk Schr\"odinger equation with energy $\varepsilon$ and quasimomentum $k_x$.

This wave function differs from the zigzag edge state wave function: first, our $\Psi$ has non-zero eigenenergy, second, it is a sum of two terms with unequal localization length, while a zigzag edge state is characterized by a single decay length.

\section{Edge with modified hopping integral}
%%%%%%%%%%%%%%%%%%%%%%%%%%%%%%%%%%%%%%%%%%%%%%%%%%
\label{section_dt}
%%%%%%%%%%%%%%%%%%%%%%%%%%%%%%%%%%%%%%%%%%%%%%%%%%

\begin{figure}
\includegraphics[width=\columnwidth]{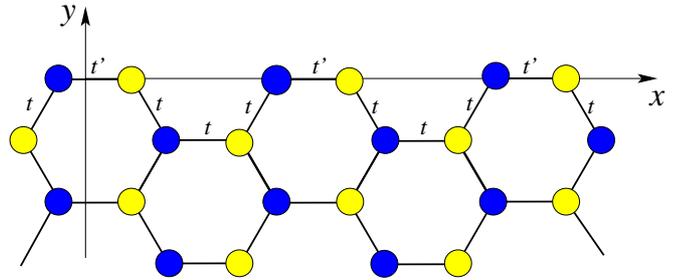}
\caption{(Color online) The model studied in Sec. \ref{section_dt}. Graphene occupies $y \leq 0$ half-plane. Nearest-neighbor hopping integral $t'$ between atoms at the edge is different from the hopping integral $t$ in the bulk. Localized states propagate along the $x$-axis and decay for $y\rightarrow -\infty$.}
\label{fig2armchair}
\end{figure}

In the previous section we derived two most general forms of the decaying wave function, Eqs.~\eqref{wf2} and~\eqref{wf1}, which satisfy the bulk Schr\"odinger equation. A priori, however, these wave functions are not necessary consistent with the Schr\"odinger equation near the edge. Below we will investigate under what conditions these wave functions satisfy the Schr\"odinger equation near the edge.

If electronic behavior near the armchair edge is described by the nearest-neighbor hopping Hamiltonian with constant hopping amplitude, no localized states exist \cite{no_edge_state_armchair2012}. To localize electrons at the edge this model has to be modified. In this Section we study a particular modification which stabilizes the edge state band. Namely, we will assume that hopping integral between carbon atoms at the edge, $t'$, is different from $t$, the hopping integral in the bulk (see Fig. \ref{fig2armchair}): $t' \neq t$.

Under such an assumption the Schr\"odinger equation for atoms at the edge
($y=0$) may be written as:
\begin{equation}
\begin{array}{r}
\varepsilon a(x,0) = -t' b(x,0) -t b(x-\frac{3a_0}{2},-\frac{\sqrt 3 a_0}{2}), \\
\\
\varepsilon b(x,0) = -t' a(x,0) -t a(x+\frac{3a_0}{2},-\frac{\sqrt 3 a_0}{2}).
\end{array}
\label{bc6}
\end{equation}
This system of equations differs from the bulk Schr\"odinger equation ~\eqref{schrodmain}. It acts as a boundary condition for the wave function of electrons.

\subsection{Type A solution}

Here we search for a localized-state eigenfunction in the form given by Eq.~\eqref{wf2}. By construction, such a wave function satisfies the Schr\"odinger equation in the bulk.

We must choose coefficients $C_{1,2}$ in such a manner that the boundary condition, Eq.~\eqref{bc6}, is also satisfied. Substituting $\Psi$ into Eq.~\eqref{bc6} we obtain:
\begin{equation}
\begin{array}{r}
\varepsilon(C_1 \alpha_1+C_2 \alpha_2)= -t'(C_1 \beta_1 + C_2 \beta_2) -  \\
-t e^{-i\bar{k}_x}(C_1 \beta_1 e^{-i\bar{k}_y-\bar{\varkappa}} +C_2 \beta_2 e^{i\bar{k}_y-\bar{\varkappa}}),\\
\\
\varepsilon(C_1 \beta_1+C_2 \beta_2)= -t'(C_1 \alpha_1 + C_2 \alpha_2) -\\
-t e^{i\bar{k}_x}(C_1 \alpha_1 e^{-i\bar{k}_y-\bar{\varkappa}} +C_2 \alpha_2 e^{i\bar{k}_y-\bar{\varkappa}}).
\end{array}
\label{bcm1}
\end{equation}
This is a system of linear equations for $C_1 $ and $C_2$. It has non-trivial solutions only when its determinant is zero:
\begin{equation}
\left|
\begin{array}{cc}
A_{11} & A_{12}\\
A_{21} & A_{22}
\end{array}\right|=0,
\label{det1}
\end{equation}
where components $A_{ij}$ are
\begin{equation}
\begin{array}{l}
A_{11}=\varepsilon \alpha_1 +t'\beta_1 +t\beta_1 e^{-i\bar{k}_x -i\bar{k}_y-\bar{\varkappa}},\\
A_{12}=\varepsilon \alpha_2 +t'\beta_2 +t\beta_2 e^{-i\bar{k}_x +i\bar{k}_y-\bar{\varkappa}},\\
A_{21}=\varepsilon \beta_1 +t'\alpha_1 +t\alpha_1 e^{i\bar{k}_x -i\bar{k}_y-\bar{\varkappa}},\\
A_{22}=\varepsilon \beta_2 +t'\alpha_2 +t\alpha_2 e^{i\bar{k}_x +i\bar{k}_y-\bar{\varkappa}}.
\end{array}
\label{aij1}
\end{equation}
It is convenient to simplify Eqs.~\eqref{aij1} with the help of the bulk Schr\"odinger equation \eqref{schac}:
\begin{equation}
\begin{array}{l}
A_{11}=(t'-t)\beta_1 -t\beta_1 e^{-i\bar{k}_x +i\bar{k}_y+\bar{\varkappa}},\\
A_{12}=(t'-t)\beta_2 -t\beta_2 e^{-i\bar{k}_x -i\bar{k}_y+\bar{\varkappa}},\\
A_{21}=(t'-t)\alpha_1 -t\alpha_1 e^{i\bar{k}_x +i\bar{k}_y+\bar{\varkappa}},\\
A_{22}=(t'-t)\alpha_2 -t\alpha_2 e^{i\bar{k}_x -i\bar{k}_y+\bar{\varkappa}}.
\end{array}
\label{aij2}
\end{equation}
Using these expressions we can evaluate determinant in Eq.~\eqref{det1} and obtain the following equation:
%\begin{equation}\label{condition1}
%\sinh  \bar{\varkappa}   \left[ \delta \tau ^2+e^{2 \bar{\varkappa}}- \delta \tau e^{  \bar{\varkappa} }\frac{\cos^2  \bar{k}_x }{\cosh\bar{\varkappa} }  \right]=\delta \tau e^{ \bar{\varkappa} } \sin^2 \bar{k}_x\,,
%\end{equation}
\begin{equation}\label{condition1}
\frac{\delta\tau}{e^{\bar{\varkappa}}}+\frac{e^{\bar{\varkappa}}}{\delta\tau}=
\frac{\cos^2\bar{k}_x}{\cosh\bar{\varkappa}}+\frac{\sin^2\bar{k}_x}{\sinh\bar{\varkappa}}\,,
\end{equation}
where
\begin{equation}
\delta \tau= \frac{t-t'}{t}.
\end{equation}
Deriving this equation we used the following relations between $\alpha$ and $\beta$ [see Eq.~\eqref{pq}]:
\begin{equation}
\frac{\alpha_1}{\beta_1}=\frac{p_1}{\varepsilon},~~\frac{\alpha_2}{\beta_2}=\frac{p_2}{\varepsilon},
\end{equation}
where
\begin{equation}
p_{1,2}=-ite^{-i\bar{k}_x}\left(\sin \bar{k}_x \pm 2\sin \bar{k}_y\sinh \bar{\varkappa}\right)
\label{p12}
\end{equation}

\begin{figure}
\includegraphics[width=1\linewidth]{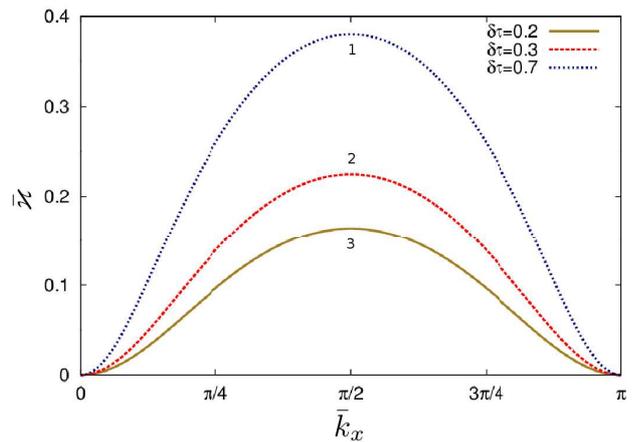}
\caption{(Color online) Numerical solution of Eq.~\eqref{condition1} gives $\bar{\varkappa}$ as a function of $\bar{k}_x$ for different $t'$: (1) $\delta \tau =0.7$, (2) $\delta \tau=0.3$, (3) $\delta  \tau =0.2$. One can observe that the closer $t'$ to $t$, the closer $\bar{\varkappa}$ to zero. If $t'=t$, then $\bar{\varkappa}=0$. This means that on unmodified armchair edge the localized states do not exist.}
\label{kappa1}
\end{figure}

Equation \eqref{p12} is equivalent to Eq.~\eqref{pq} for type A wave function. The use of Eq.~\eqref{p12} significantly simplifies the derivation of Eq.~\eqref{condition1}.

\begin{figure}
\includegraphics[width=1\linewidth]{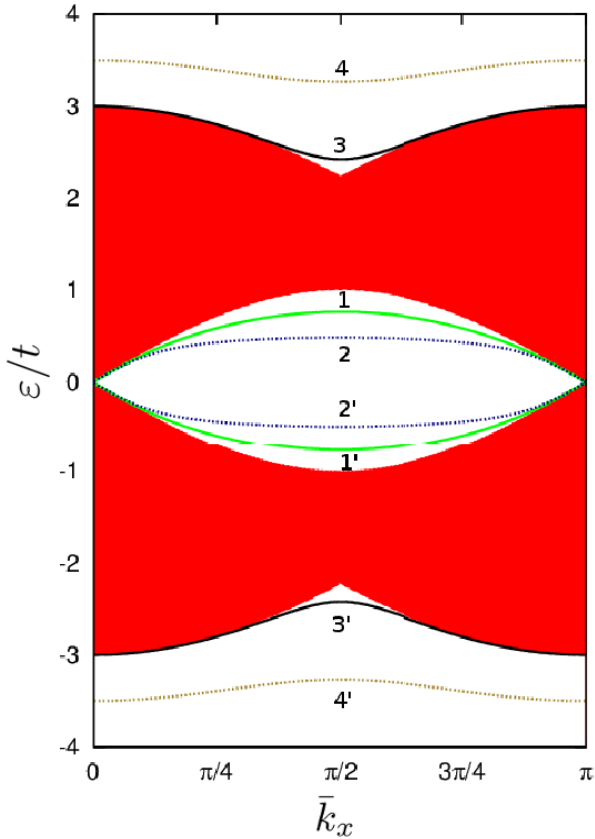}
\caption{(Color online) Armchair edge states and the bulk states of
graphene. For given $k_x$ there is a continuum of the graphene bulk states,
corresponding to different $k_y$. This continuum is shown as a red shaded
area. The curves $1-4$ and $1'-4'$ represent the bands of edge states,
calculated numerically for different $t'$. For $t'<t$ electron-hole
symmetrical pairs of edge bands exist in the gap between the continuum of
bulk states: (1-1') $t'=0.5t$, (2-2') $t'=0.1t$. These are type A
solutions. When $t'>2t$ localized states emerge above and below the bulk
states: (3-3') $t'=2.01t$, (4-4') $t'=2.5t$. These are type B solutions.}
\label{p2energy}
\end{figure}

Equation \eqref{condition1}, together with Eqs.~\eqref{ime2} and \eqref{ree2}, defines $\varepsilon$, $k_y$, and  $\varkappa$ as implicit functions of $k_x$. Numerical solution to Eq.~\eqref{condition1} for $\bar{\varkappa}(\bar{k}_x)$ is plotted in Fig.~\ref{kappa1} for different values of $t'$, while results for $\varepsilon(\bar{k}_x)$ are shown in Fig.~\ref{p2energy}.

Let us now discuss the derived results. An important piece of information can be easily obtained from Eq.~\eqref{condition1}. Its right hand side is always positive. Thus, if $\delta \tau <0$, no solution of Eq.~\eqref{condition1} exists. This means that type A edge states may be found only when $t'<t$.

In addition to numerical results, we obtain approximate expressions for $\varepsilon,\bar{\varkappa},\bar{k}_y$ in two limits.
First, we can calculate these functions close to Dirac cone ($\bar{k}_x \rightarrow 0$, $\bar{k}_y \rightarrow \frac{2 \pi}{3}$), that is near the Fermi level of the graphene:

\begin{eqnarray}
\bar{\varkappa}&\cong& \frac{\delta \tau \bar{k}_{x}^{2}}{\delta \tau^2+1-\delta \tau},\\
\bar{k}_y&\cong & \frac{2\pi}{3}-\frac{\bar{k}_x^2}{2\sqrt3} ,\\
\varepsilon/t& \cong & \bar{k}_x-\left(\frac{3 \delta \tau^2}{2\left(1-\delta \tau+\delta \tau^2\right)^2}+\frac{1}{6}\right)\bar{k}_x^3. \label{appEkxformula}
\end{eqnarray}

\begin{figure}
\includegraphics[width=1\linewidth]{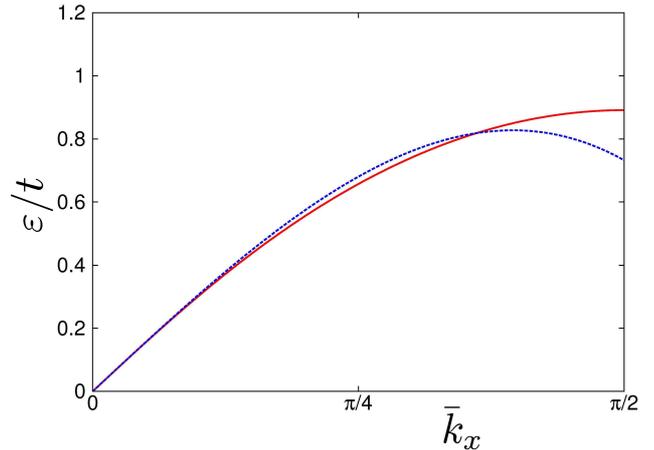}
\caption{(Color online) Eigenenergy $\varepsilon$ as function of $\bar{k}_x$ for $\delta\tau=0.3$. Solid (red) curve shows the numerical solution of Eq.~\eqref{condition1}. Dotted (blue) curve corresponds to the approximate formula~\eqref{appEkxformula}, which is valid for small $\bar{k}_x$. We can see that Eq.~\eqref{appEkxformula} is in a good agreement with the numerical result for $k_x \lesssim \pi/4$.}
\label{appkx}
\end{figure}

Accuracy of Eq.~\eqref{appEkxformula} can be estimated by examining Fig.~\ref{appkx}. We see that for $\delta\tau=0.3$, Eq.~\eqref{appEkxformula} for $\varepsilon$ works well for $|k_x|\lesssim\pi/4$.

Second, we study Eq.~\eqref{condition1} in the limit of small deformations: $\delta \tau \ll 1$. Functions of $\varepsilon(\bar{k}_x),\bar{\varkappa}(\bar{k}_x),\bar{k}_y(\bar{k}_x)$ are approximated by:
\begin{align}
\bar{\varkappa} \cong {}& \delta \tau \sin^2 \bar{k}_x,\\
\bar{k}_y \cong {}& \arccos\left(-\frac{\cos \bar{k}_x}{2}\right) -\frac{\sin^4 \bar{k}_x \cos \bar{k}_x}{4\sqrt{1-\frac{\cos^2 \bar{k}_x}{4}}}\delta \tau^2, \label{kyapp} \\
\varepsilon/t \cong {}& \sin \bar{k}_x - 2\delta \tau^2 \sin^3 \bar{k}_x\left(1-\frac{\cos^2 \bar{k}_x}{4}\right). \label{appEdtformula}
\end{align}

\begin{figure}
\includegraphics[width=1\linewidth]{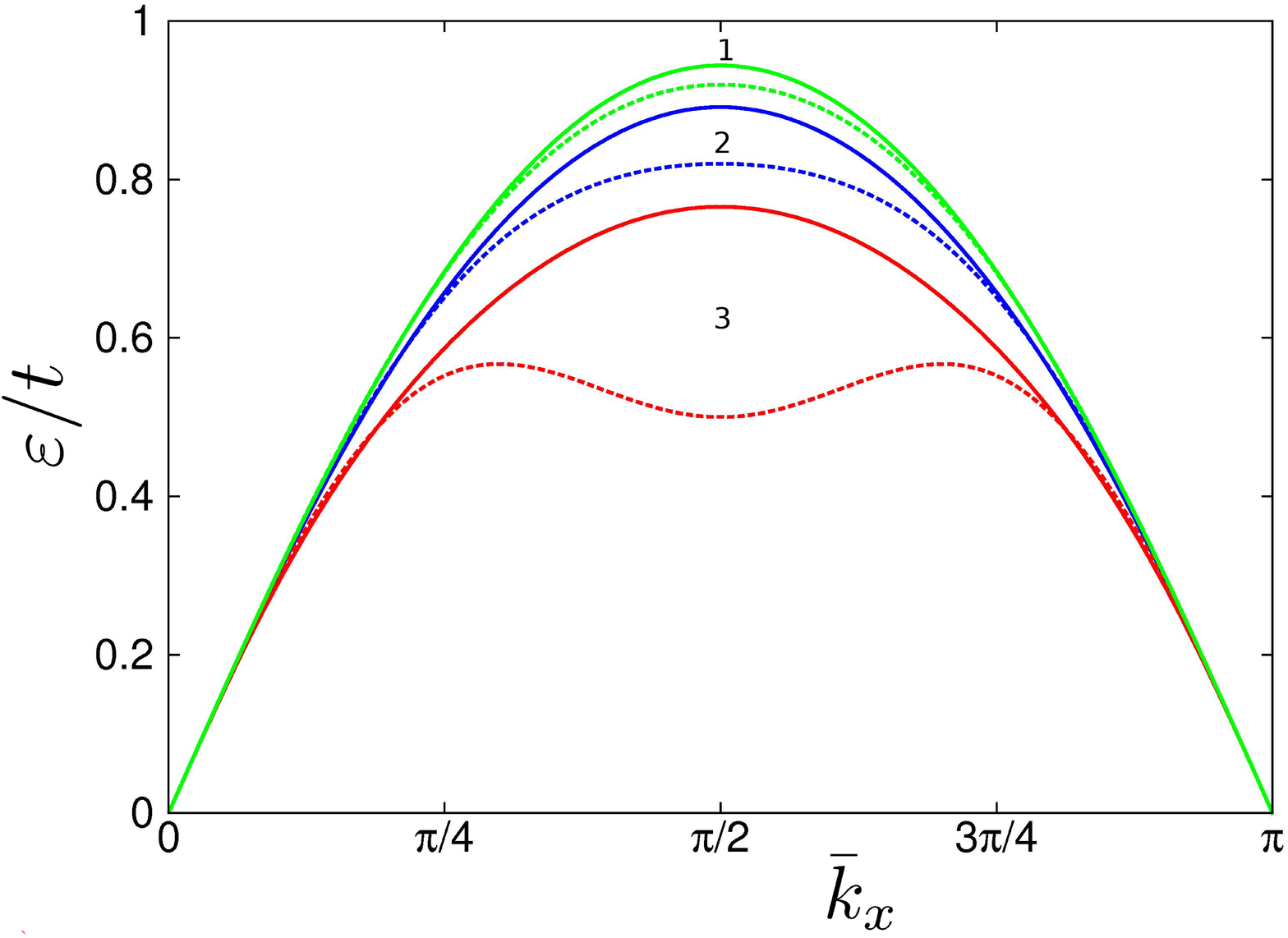}
\caption{(Color online) The function $\varepsilon(\bar{k}_x)$ calculated numerically (solid curves) and using approximation~\eqref{appEdtformula} in the limit of $\delta\tau \ll 1$: (1) $\delta \tau =0.2$, (2) $\delta \tau =0.3$,(3) $\delta \tau =0.5$. Even for $\delta \tau$ as large as 0.3 Eq.~\eqref{appEdtformula} works very well.}
\label{appdt}
\end{figure}

To estimate the accuracy of this approximation for different $\delta\tau$ we plot $\varepsilon(\bar{k}_x)$ calculated numerically together with Eq.~\eqref{appEdtformula} in Fig. \ref{appdt}. We see that Eq.~\eqref{appEdtformula} is accurate for $\delta \tau \lesssim 0.3$.

It is interesting that $\bar{k}_y$ demonstrates a very weak dependence on $t'$. Indeed, Eq.~\eqref{kyapp} does not contain a term linear in $\delta\tau$. In addition, the factor before $\delta \tau^2$ is very small for all $k_x$ (its maximum value is about $0.1$). As a result, the first term in Eq.~\eqref{kyapp} approximates the dependence $\bar{k}_y(\bar{k}_x)$. Non-zero $\bar{k}_y$ leads to oscillations of the edge state electron density with the $y$-coordinate, which, in principle, can be observed experimentally.

The results of numerical calculations of $\varepsilon(\bar{k}_x)$ for two
different values of $t'$ are shown in Fig. \ref{p2energy}. The type A edge
band $\varepsilon(k_x)$ is surrounded by the continuum of the bulk graphene
states (also shown in this figure), and for given $t'$ there are two
particle-hole symmetrical edge bands $\varepsilon_{1,2}(\bar{k}_x)$, such
that $\varepsilon_{2}(\bar{k}_x)=-\varepsilon_{1}(\bar{k}_x)$.

\subsection{Type B solution}

Next, we discuss the type B solution, Eq.~\eqref{wf1}.

Substituting wave function $\Psi$ into the boundary conditions Eq.~\eqref{bc6} we obtain:
\begin{align}
\varepsilon (C_1 \alpha_1 &+ C_2 \alpha_2 ) = -t'(C_1 \beta_1 + C_2 \beta_2 ) - \nonumber \\
&-t \left(C_1 \beta_1 e^{-i\bar{k}_x-\bar{\varkappa}_1} - C_2 \beta_2 e^{-i\bar{k}_x-\bar{\varkappa}_2} \right),
\end{align}
\begin{align}
\varepsilon (C_1 \beta_1 &+ C_2 \beta_2 ) = -t'(C_1 \alpha_1 + C_2 \alpha_2 ) - \nonumber \\
&-t \left( C_1 \alpha_1 e^{i\bar{k}_x-\bar{\varkappa}_1}-C_2 \alpha_2 e^{i\bar{k}_x-\bar{\varkappa}_2} \right).
\end{align}
The system of linear equations for $C_1$, $C_2$ has non-trivial solutions only if its determinant is zero. This occurs when the following condition holds true:
\begin{align}
( \cosh \bar{\varkappa}_1 &+ \cosh \bar{\varkappa}_2 )\!\!\left[ \delta \tau^2-\delta \tau \cos \bar{k}_x ( e^{\bar{\varkappa}_2}-e^{\bar{\varkappa}_1}) -e^{\bar{\varkappa}_1+\bar{\varkappa}_2}\right]\nonumber \\
&-\delta \tau \sin^2 \bar{k}_x (e^{\bar{\varkappa}_1}+e^{\bar{\varkappa}_2})=0.
\label{p2sinequation}
\end{align}
The derivation of this equation is similar to derivation of Eq.~\eqref{condition1}.

\begin{figure}
\includegraphics[width=1\linewidth]{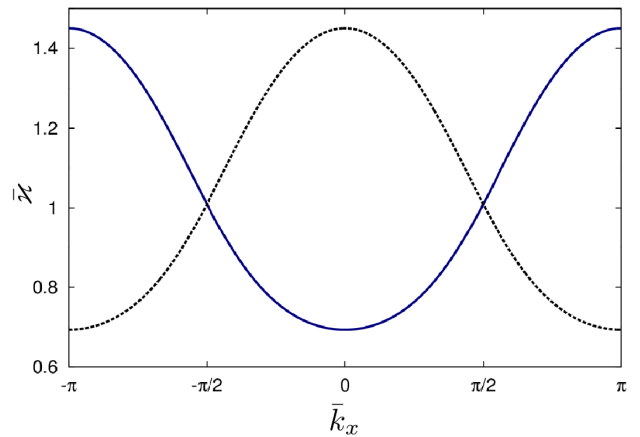}
\caption{(Color online) Functions $\bar{\varkappa}_1(\bar{k}_x)$ (solid curve) and $\bar{\varkappa}_2(\bar{k}_x)$ (dotted curve) found by solving Eq.~\eqref{p2sinequation} (type B solution) . The parameter $t'=3t$. One can see that $\bar{\varkappa}_1(\bar{k}_x)=\bar{\varkappa}_2(\bar{k}_x+\pi)$.}
\label{p2kappa}
\end{figure}

The system of Eqs.~\eqref{energyB}, \eqref{condkapB}, and~\eqref{p2sinequation} must be solved to find $\bar{\varkappa}_{1,2}(\bar{k}_x)$ and $\varepsilon(\bar{k}_x)$. The solution exists only when $t'>2t$. The dependencies $\bar{\varkappa}_1(\bar{k}_x)$ and $\bar{\varkappa}_2(\bar{k}_x)$ are shown in Fig. \ref{p2kappa}. These functions oscillate with $\bar{k}_x$. One can show from Eqs.~\eqref{condkapB} and \eqref{p2sinequation} that $\bar{\varkappa_2}(\bar{k}_x+\pi)=\bar{\varkappa_1}(\bar{k}_x)$. Energy as a function of $\bar{k}_x$ is plotted in Fig. \ref{p2energy}. The localized-state bands lie symmetrically above and below the continuum of the bulk states. In the range
\begin{equation}
t<t'<2t
\end{equation}
localized states do not exist.

\section{Functionalized armchair edge}
%%%%%%%%%%%%%%%%%%%%%%%%%%%%%%%%%%%%%%%%%%%%%%%%%%
\label{section_radical}
%%%%%%%%%%%%%%%%%%%%%%%%%%%%%%%%%%%%%%%%%%%%%%%%%%

\begin{figure}
\includegraphics[width=0.95\columnwidth]{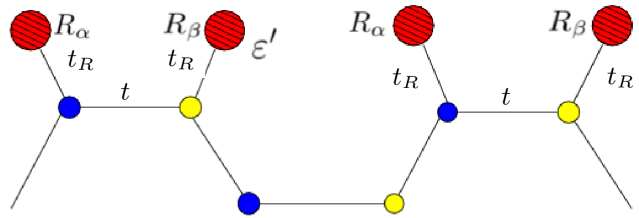}
\caption{(Color online) Graphene armchair edge with radical atoms attached. Here $R_{\alpha,\beta}$ are wave functions of electrons at the radical sites, $\varepsilon'$ is on-site potential, $t_R$ is a hopping integral between radicals and nearest carbon atoms, which may be different from $t$, hopping integral in graphene sheet.}
\label{armchairfig2}
\end{figure}

It is often assumed in the theoretical literature, that hydrogen atoms or
monovalent radicals of some other type are attached to the edge to
saturate dangling $sp^2$-bond at the edge. Here we would like to discuss a
more complicated situation. We will assume that, in addition to the
formation of the bond with $sp^2$ electrons, the attached radicals have an
extra orbital which hybridizes with a $\pi$-orbital of carbon. If the physics of $\pi$-electrons is discussed, these
orbitals have to be accounted for: they appear as additional
sites where $\pi$-electrons can hop to (see
Fig.~\ref{armchairfig2}).
We will demonstrate that such `functionalized armchair edge' also supports localized states.

The boundary condition for electrons near the functionalized edge is:
\begin{equation}
\label{boundcondrad}
\begin{array}{rcl}
\varepsilon a(x,0)\!&=&\!-t b(x,0) -t b\left(x-\frac{3a_0}{2},-\frac{\sqrt3 a_0}{2}\right)-t_R R_{\alpha}(x),\\
\varepsilon b(x,0)\!&=&\!-t a(x,0) -t a\left(x+\frac{3a_0}{2},-\frac{\sqrt3 a_0}{2}\right)-t_R R_{\beta}(x),\\
\varepsilon R_{\alpha}(x)\!&=&\!\varepsilon'R_{\alpha}(x)-t_R a(x,0),\\
\varepsilon R_{\beta}(x)\!&=&\!\varepsilon'R_{\beta}(x)-t_R b(x,0),
\end{array}
\end{equation}
where $R_{\alpha,\beta}(x)$ are wave functions of electrons at the radical sites (see Fig.~\ref{armchairfig2}), $\varepsilon'$ is on-site potential for radical sites, and $t_R$ is a hopping integral between radicals and nearest carbon atoms at the edge. Excluding $R_{\alpha,\beta}(x)$ from these equations, one obtains the boundary condition for the electron's wave function in graphene:
\begin{equation}\label{boundcondrad2}
\begin{array}{rcl}
\left(\varepsilon-\frac{t_R^2}{\varepsilon-\varepsilon'}\right)a(x,0)\!&=&\!-t b(x,0)-t b\left(x-\frac{3a_0}{2},-\frac{\sqrt3 a_0}{2}\right),\\
\\
\left(\varepsilon-\frac{t_R^2}{\varepsilon-\varepsilon'}\right)b(x,0)\!&=&\!-t a(x,0)-t a\left(x+\frac{3a_0}{2},-\frac{\sqrt3 a_0}{2}\right).
\end{array}
\end{equation}

Using these equations, we perform the analysis of the edge states similar to that done in the previous section.

\subsection{Type A solution}

For the type A solution, the wave function has a form of Eq.~\eqref{wf2}. Substituting this expression into Eq.~\eqref{boundcondrad2}, we obtain the following system of equations for the coefficients $C_{1,2}$:
\begin{align}
(C_1 \alpha_1 + C_2 \alpha_2)\left(\varepsilon - \frac{t_R^2}{\varepsilon - \varepsilon '}\right)=-t(C_1 \beta_1 + C_2 \beta_2) - \nonumber \\
-t\left(C_1 \beta_1 e^{-i\bar{k}_x} e^{-i\bar{k}_y-\bar{\varkappa}}+C_2 \beta_2 e^{-i\bar{k}_x} e^{i\bar{k}_y-\bar{\varkappa}}\right),\label{linc1}\\
(C_1 \beta_1 + C_2 \beta_2)\left(\varepsilon - \frac{t_R^2}{\varepsilon - \varepsilon '}\right)=-t(C_1 \alpha_1 + C_2 \alpha_2) - \nonumber \\
-t\left(C_1 \alpha_1 e^{i\bar{k}_x} e^{-i\bar{k}_y-\bar{\varkappa}}+C_2 \alpha_2 e^{i\bar{k}_x} e^{i\bar{k}_y-\bar{\varkappa}}\right).\label{linc2}
\end{align}
This system has non-trivial solutions, if the determinant of the following matrix is zero:
\begin{equation}
\left|\begin{array}{cc}
M_{11} & M_{12}\\
M_{21} & M_{22}
\end{array}\right|=0,
\label{determinant}
\end{equation}
where
\begin{equation}
\begin{array}{l}
M_{11}=\alpha_1\left(\varepsilon-\frac{t_R^2}{\varepsilon - \varepsilon'}\right)+t\beta_1+t \beta_1 e^{-i\bar{k}_x} e^{-i\bar{k}_y-\bar{\varkappa}},\\
M_{12}=\alpha_2 \left(\varepsilon-\frac{t_R^2}{\varepsilon - \varepsilon'}\right)+t \beta_2+t \beta_2 e^{-i\bar{k}_x} e^{i\bar{k}_y-\bar{\varkappa}},\\
M_{21}=\beta_1 \left(\varepsilon-\frac{t_R^2}{\varepsilon - \varepsilon'}\right)+t \alpha_1+t \alpha_1 e^{i\bar{k}_x} e^{-i\bar{k}_y-\bar{\varkappa}},\\
M_{22}=\beta_2\left(\varepsilon-\frac{t_R^2}{\varepsilon - \varepsilon'}\right)+t \alpha_2+t \alpha_2 e^{i\bar{k}_x} e^{i\bar{k}_y-\bar{\varkappa}}.
\end{array}
\end{equation}
After straightforward algebra we derive equation for $\bar{\varkappa}$:
\begin{equation}\label{p4eq1}
\varepsilon e^{\bar{\varkappa}} \tau_R^2 (\varepsilon-\varepsilon')=\sinh\bar{\varkappa}\left[\tau_R^4t^2-e^{2\bar{\varkappa}}(\varepsilon-\varepsilon')^2\right],
\end{equation}
where
\begin{equation}
\tau_R=\frac{t_R}{t}\,.
\end{equation}
Let us remember that for the type A solution, the energy $\varepsilon$ as a
function of $\bar{\varkappa}$ and $\bar{k}_x$ is given by Eq.~\eqref{ree2}.
Solving Eq.~\eqref{p4eq1} together with Eq.~\eqref{ree2}, we find the
spectrum of the edge band
$\varepsilon(\bar{k}_x)$.

Properties of this set of the equations depend on
$t_R/t$
and
$\varepsilon'/t$.
Specifically, the number of the edge-states branches varies as these
parameters change: for some parameter values no edge states exist,
for others as many as six branches are present. In this section we will
study the limit of large
$t_R \gg t$.
Other regimes are discussed in
%limiting cases: {\it (1)} large $t_R\gg t$, {\it (2)} small $t_R\ll t$ and
%{\it (3)} large $|\varepsilon'|\gg t,t_R$. Case {\it (1)} is studied in
%this chapter, another two cases are studied in the
Appendix~\ref{appendixA}.

\begin{figure}[H]
\includegraphics[width=\columnwidth]{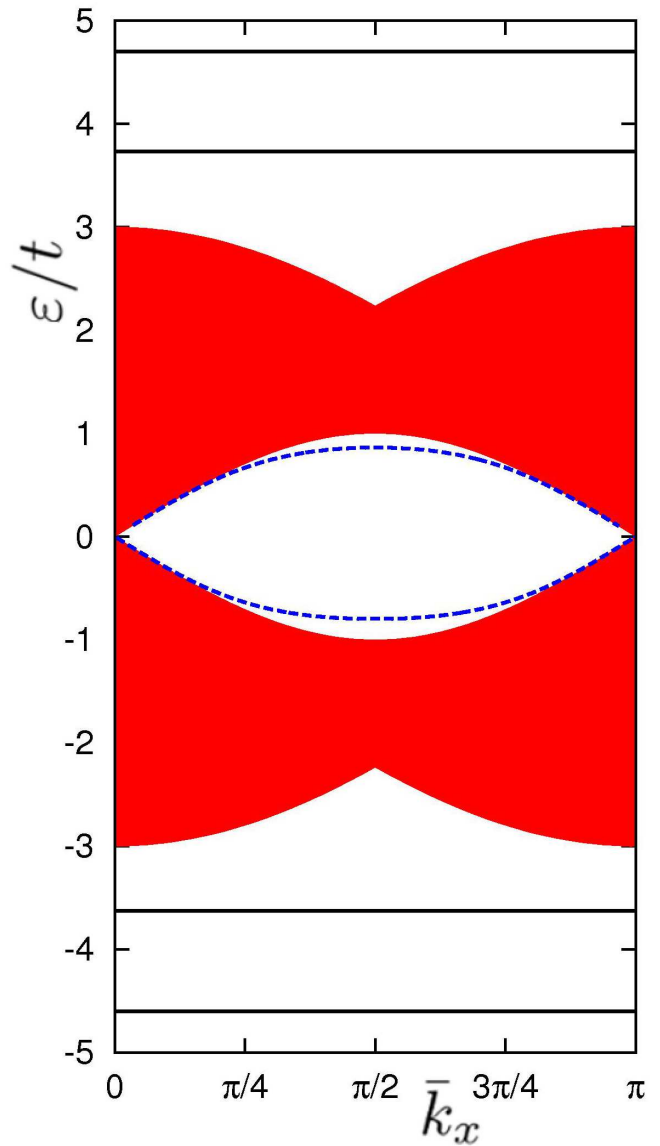}
\caption{
(Color online) Energy of localized electron states near the
functionalized armchair edge calculated for large
$t_R \gg t$.
There are two edge bands
$\varepsilon_{1,2}(k_x)$
for type A solution (blue, dashed) found numerically for
$\varepsilon'=0.1t$,
$t_R=2t$. For such a choice of the parameters type B solutions are absent.
For same
$\varepsilon'=0.1t$,
but larger
$t_R = 4 t$
there are four non-dispersive edge bands of type B (black solid lines). In addition, two type A solutions exist. However, since their eigenenergies
lie very close to the edge of the continuum, these branches are not shown.
The shaded area (red) corresponds to the bulk graphene states.
}
\label{p42en}
\end{figure}

If
$t_R \gg t$,
at least one solution of
Eq.~\eqref{p4eq1}
exists for any $\varepsilon'$. It follows from
Eq.~\eqref{p4eq1}
that for large
$t_R$
the inverse localization length
$\varkappa$
is small. Thus, the type A wave functions spread deeply into the bulk in
this regime. From
Eq.~\eqref{p4eq1}
one derives
\begin{eqnarray}
%%%%%%%%%%%%%%%%%%%%%%%%%%%%%%%%%%%%%%%%%%%%%%%%%%
\label{kappa_A}
%%%%%%%%%%%%%%%%%%%%%%%%%%%%%%%%%%%%%%%%%%%%%%%%%%
\bar{\varkappa}
\cong
\frac{1}{\tau_R^2}
|\sin(\bar{k}_x)|
\left(
	|\sin(\bar{k}_x)|\pm\frac{|\varepsilon'|}{t}
\right)\,,
\\
%%%%%%%%%%%%%%%%%%%%%%%%%%%%%%%%%%%%%%%%%%%%%%%%%%
\label{eigenenergy_A}
%%%%%%%%%%%%%%%%%%%%%%%%%%%%%%%%%%%%%%%%%%%%%%%%%%
\varepsilon(\bar{k}_x)
=
\pm\text{sgn}(\varepsilon')t|\sin(\bar{k}_x)|+O(\bar{\varkappa}^2),
\end{eqnarray}
where two signs correspond to two branches of the edge states. These
branches are located near the edges of the bulk continuum (see
Fig.~\ref{p42en}).
Since
$\bar\varkappa$
must be positive, the solution corresponding to minus sign in
Eqs.~(\ref{kappa_A})
and~(\ref{eigenenergy_A})
exists only when
$|\sin(\bar{k}_x)|>|\varepsilon'|/t$,
and completely disappears for
$|\varepsilon'|>t$.

\subsection{Type B solution}

Substituting type B wave function,
Eq.~\eqref{wf1},
into the boundary
condition~\eqref{boundcondrad2},
and performing calculations similar to that described before
Eq.~\eqref{p4eq1},
we obtain the following equation:
\begin{align}
(\cosh \bar{\varkappa}_1&+\cosh\bar{\varkappa}_2)
\left[
	\tau_R^4t^2
	+
	(\varepsilon-\varepsilon')^2
	e^{\bar{\varkappa}_1+\bar{\varkappa}_2}
\right]
\nonumber
\\
&=\tau_R^2(\varepsilon  - \varepsilon')
(e^{\bar{\varkappa}_1}+e^{\bar{\varkappa}_2})\varepsilon.
\label{p4condition2}
\end{align}
Solving this equation, together with
Eqs.~\eqref{energyB}
and~\eqref{condkapB},
for $\varepsilon$ and
$\bar{\varkappa}_{1,2}$,
we obtain the spectrum of the edge band
$\varepsilon(\bar{k}_x)$.

In the case of $t_R \gg t$ there are four particle-hole non-symmetric
solutions to Eq.~\eqref{p4condition2}.
We denote these solutions
$\varepsilon_{1,2}^{\pm}(\bar{k}_x)$.
Of these four, two solutions
[$\varepsilon_{1,2}^{-}(\bar{k}_x)$]
lie below, and two solutions
[$\varepsilon_{1,2}^{+}(\bar{k}_x)$]
lie above the bulk states. The result of numerical calculations of the edge
bands is shown in
Fig.~\ref{p42en}.
It is possible to obtain analytic results for the spectra of the edge
bands in this limit. The states are strongly localized near the edge:
$\bar{\varkappa}_{1,2}\approx\ln\tau_R \to \infty$,
when
$t_R\to\infty$.
We seek a solution to
Eq.~\eqref{p4condition2}
in the form of expansion
$\bar{\varkappa}_{1,2}
=
\ln\tau_R
+
\varkappa^{(1)}_{1,2}/\tau_R
+
\varkappa^{(2)}_{1,2}/\tau_R^2+\dots$
and $\varepsilon=\pm t_R+\varepsilon^{(1)}+\varepsilon^{(2)}/\tau_R+\dots$.
As a result, we obtain
\begin{equation}\label{E12trBig}
\varepsilon_{1,2}^{s}(\bar{k}_x)=st_R+\frac{\varepsilon'\pm t}{2}+\frac{s\left[(\varepsilon'\mp t)^2+4t^2\right]}{8t_R}+O\left(\frac{1}{\tau_R^2}\right),
\end{equation}
where $s=\pm 1$.
Note that these solutions have very weak dispersion, the largest
$\bar{k}_x$-depending
term  in the expansion for $\varepsilon$ has the order
$1/\tau_R^2$.

\subsection{Number of solutions}

We already mentioned that the number of the edge branches depends on the
Hamiltonian parameters. For a particular example of this phenomenon see our
discussion of
Eq.~(\ref{eigenenergy_A}).
Here we will briefly summarize our knowledge about the number of the
branches in different regions of the parameter space. Details may be found
in Appendices.

When $t_R \ll t$, there are two type A solutions if
$|\varepsilon'|\lesssim t$,
and two type B solutions if
$|\varepsilon'|\gtrsim \sqrt{5}t$. In the range
$t\lesssim|\varepsilon'|\lesssim \sqrt{5}t$, no solutions exist. For
$|\varepsilon'|$ very close either to $t$ or to $\sqrt{5}t$ only one
solution of corresponding type exists. All solutions in this limit have
weak dispersion.

The largest number of edge bands exists in the opposite limit
$t_R \gg t$.
In this case there are six solutions ($4$ solutions of type B and
$2$ solutions of type A) when $|\varepsilon'|\lesssim t$ or five solutions
($4$ of type B and $1$ of type A) if $|\varepsilon'|\gtrsim t$. Type A
solutions have pronounced dispersion, while all type B solutions are almost
non-dispersive.

When $|\varepsilon'|\gg t,t_R$, there are three edge bands: one of type A
and two of type B. Similar to the case $t_R \gg t$, solution of type A has
pronounced dispersion, while type B solutions have weak dispersion.

\section{Discussion}\label{discussion}

\begin{figure}
\includegraphics[width=\columnwidth]{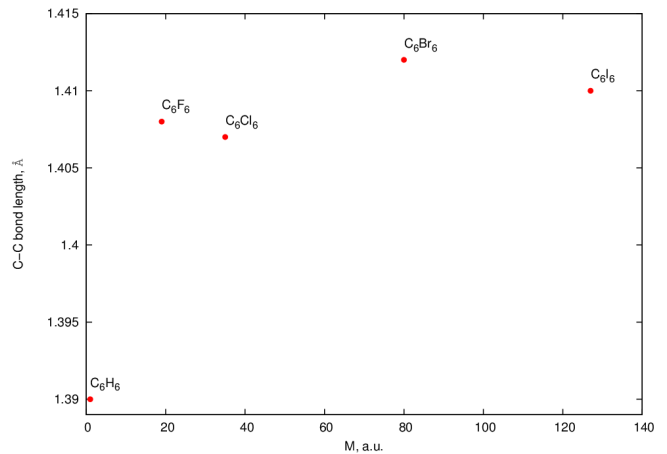}
\caption{(Color online) The length of C-C bond in benzene-like molecule
C$_6$Ha$_6$,
where Ha denotes halogen atoms, as a function of the atomic weight of
halogen atom. The weight is given in atomic units, a.u., the bond length is
in angstr\"oms. Halogen substitution leads to elongation of the C-C bond
length as compared to the bond length in an ordinary benzene molecule. The
dependence illustrates the fact that suitably chosen functionalization may,
in principle, induce sufficient elongation of the edge bond to stabilize
the edge states described in
Sec.~\ref{section_dt}.
Based on the data provided by NIST Chemistry WebBook.
}\label{cclength}
\end{figure}

We demonstrate that, as a result of the boundary conditions modification,
the localized states at the armchair graphene edge become stable in a wide
range of the parameter space. Two possible modifications are considered:
{\it (i)} the hopping integrals between the nearest-neighbor carbon atoms
near the edge $t'$ are different from that in the bulk $t$, and
{\it (ii)} non-carbon atoms attached to passivate dangling $sp^2$-bonds
also have orbitals which hybridize with the $\pi$-electrons from graphene.

Depending on the type of edge modification, {\it (i)} or {\it (ii)}, properties
of the emergent localized band differ. Namely, if the hopping integral at
the edge is modified [case {\it (i)}], the eigenenergy of the localized
states has pronounced dependence on the electron momentum (see
Fig.~\ref{p2energy}).
At the same time, when graphene $\pi$-orbitals hybridize with the
non-carbon orbitals near the edge [case {\it (ii)}], the resultant bands may
be nearly flat (see Fig.~\ref{p4energy}).
It happens, for example, when $t_R\ll t$ (see the Appendix) or $t_R\gg t$ (the type B solution, see Eq.~\eqref{E12trBig} and Fig.~\ref{p42en}). In this case, the armchair edge bands become similar to zigzag one, even though the energies of the nearly localized bands are different from zero.

The modification of the first kind, {\it (i)}, leads to the localized state
only when $t'<t$ or $t'>2t$. Due to high rigidity of the aromatic bond, it
is unlikely that $t'$ could exceed $2t$.
Can we reach the regime
$t'<t$?
{\it Ab initio}
calculations~\cite{son_gap}
show that, when the armchair termination is
%the C-C bond length about 10$\%$ less
%than a bond length in the bulk~\cite{bond_length_1,bond_length_2}, which
%leads to the hopping integral slightly greater than one in the bulk. When
passivated by hydrogen atoms, carbon-carbon bond length at the edge
is about 3.5$\%$ less than the length in the bulk. This leads to the
increase in hopping integral
$t'=1.12t>t$,
which violates the required condition. However, the length of the bond
may be altered by changing the passivating radical. For example, chemical
data show (see
Fig.~\ref{cclength})
that in benzene C$_6$H$_6$ substitution of hydrogen with higher halogens
leads to 1.4\% increase in the C-C bond length as compared to the ordinary
benzene molecule. Another
calculation~\cite{peng:023112}
shows that oxygen atoms attached to the armchair edge can lead to the
elongation of the C-C bond. We do not imply that the halogen passivation or
oxidation brings the edge into the regime of interest. Rather, these
examples demonstrate that the bond length, despite high rigidity, could be
varied to one's needs by suitable choice of passivating radical.

To create the edge modification of type
{\it (ii)} (non-carbon radicals attached to the edge,
Sec.~\ref{section_radical}),
divalent radical able to passivate the dangling
$sp^2$-bond
and to hybridize with $\pi$-electron may be suitable. Density-functional
calculations suggest that monovalent radical OH, when attached to the edge,
may act as an extra site for
$\pi$-electrons (see Fig.~7 of
Ref.~\onlinecite{Rosenkranz}
and corresponding discussion). Molecules
%NH$_2$~\cite{cervantes_func_mag},
NO$_2$,
CO$_2$,
and
O$_2$,
adsorbed on the armchair edge, demonstrate similar
behavior~\cite{huang_hybrid}.

The edge states near the modified armchair termination were studied
numerically in several
papers~\cite{li_t0,changwon,klos}.
Specifically, the model with modified hopping (similar to the model of
Sec.~\ref{section_dt})
has been discussed in
Ref.~\onlinecite{li_t0}.
This paper reports the existence of the edge band with dispersion similar
to the dispersion of our type A states (see
Fig.~\ref{p2energy}
above). However, type B solutions were not mentioned there. The edge
modification similar to our
Sec.~\ref{section_radical}
have been studied in
Ref.~\onlinecite{klos}.
In this paper solutions similar to our type A and type B were obtained
numerically.

We expect that localized states may be experimentally observed in scanning
tunneling spectroscopy. It was
demonstrated~\cite{enoki_2008}
that zigzag edge states are responsible for peak in local density of states
at the Fermi level (see Fig.~4 of
Ref.~\onlinecite{enoki_2008}).
We believe that edge states described in this paper will produce peaks in
the tunneling spectrum, whose intensity decays away from the edge.

In any realistic sample, some amount of disorder is
present, and the question of the stability of the found edge states toward
disorder arises. Although, the discussion of the disorder effects are
beyond the scope of this work, we would like, however, to offer the
following observations. There are two types of edge bands in our case: type
A and B. As one can explicitly see from Figs.~\ref{p2energy}
and~\ref{p42en}, edge bands of type A and some of type B overlap in the
energy domain with a continuum of the bulk states. Consequently, disorder
couples a given localized state with energy $\varepsilon$ to numerous bulk
states whose energies $E$ are close to $\varepsilon$. As a result, the
localized state with energy $\varepsilon$ and inverse localization length
$\varkappa$ becomes a resonance with finite lifetime
$\Gamma_{\varepsilon\varkappa}$. For small concentration of impurities
$n_{\text{imp}}$ this lifetime can be estimated as
\begin{equation}
%%%%%%%%%%%%%%%%%%%%%%%%%%%%%%%%%%%%%%%%%%%%%%%%%%
\label{Gamma}
%%%%%%%%%%%%%%%%%%%%%%%%%%%%%%%%%%%%%%%%%%%%%%%%%%
\Gamma_{\varepsilon\varkappa}
=
2\pi n_{\text{imp}}
V_0^2
\sum_{\mathbf{n}\alpha} \sum_{\mathbf{k}}
	\left|
		\psi^{\alpha}_{\varepsilon\varkappa}(\mathbf{n})
		\varphi^{\alpha}_{\mathbf{k}}(\mathbf{n})
	\right|^2
\delta\left(\varepsilon-\varepsilon_{\mathbf{k}}\right)\,,
\end{equation}
where $V_0$ is the strength of the point-like interaction between electrons
and impurity, $\varepsilon_{\mathbf{k}}$ is the energy spectrum of the bulk
electrons given by Eq.~\eqref{epsilon} with $\varkappa=0$, and
$\psi^{\alpha}_{\varepsilon\varkappa}(\mathbf{n})$ and
$\varphi^{\alpha}_{\mathbf{k}}(\mathbf{n})$ are the spinor wave functions
($\alpha={\it A,B}$) of the edge and bulk electrons, respectively. It is
assumed that impurities are distributed randomly and uniformly over the
sample. The summation over elementary unit cells
$\mathbf{n}$
of the sample, and the summation over $\mathbf{k}$ ($k_y>0$) are performed.
The edge state wave function
$\psi_{\varepsilon\varkappa}(\mathbf{n})$ is given either by
Eq.~\eqref{wf2} or by Eq.~\eqref{wf1} depending on the type of solution.
The bulk state wave function
$\varphi_{\mathbf{k}}(\mathbf{n})$
is the superposition of incident and reflected plane waves with momenta
${\bf k} = (k_x,k_y)$ and $(k_x,-k_y)$.
While we did not calculate $\varphi$ here, it can be determined with the
help of the Schr\"odinger equation
Eq.~(\ref{schrodmain})
complimented by an appropriate boundary conditions [either
Eq.~(\ref{bc6}),
or
Eq.~(\ref{boundcondrad})].

{
It is convenient to define the local density of bulk states:
\begin{eqnarray}
\rho^\alpha ({\bf n},\varepsilon)=\sum_{\mathbf{k}}	\left|\varphi^{\alpha}_{\mathbf{k}}(\mathbf{n})\right|^2\delta\left(\varepsilon-\varepsilon_{\mathbf{k}}\right)\,.
\end{eqnarray}
When averaged over the lattice, the usual density of states
$\rho (\varepsilon)$
for graphene is recovered:
\begin{eqnarray}
%%%%%%%%%%%%%%%%%%%%%%%%%%%%%%%%%%%%%%%%%%%%%%%%%%
\label{density}
%%%%%%%%%%%%%%%%%%%%%%%%%%%%%%%%%%%%%%%%%%%%%%%%%%
\frac{1}{N}
\sum_{{\bf n} \alpha}
\rho^\alpha ({\bf n}, \varepsilon)
=
\rho (\varepsilon)\,,
\end{eqnarray}
where $N$ is the number of sites in the sample. Using
$\rho^\alpha ({\bf n}, \varepsilon)$
we can re-write the expression for $\Gamma$:
\begin{equation}
%%%%%%%%%%%%%%%%%%%%%%%%%%%%%%%%%%%%%%%%%%%%%%%%%%
\label{Gamma_rho}
%%%%%%%%%%%%%%%%%%%%%%%%%%%%%%%%%%%%%%%%%%%%%%%%%%
\Gamma_{\varepsilon\varkappa}
=
2\pi n_{\text{imp}}
V_0^2
\sum_{\mathbf{n}\alpha}
	\left|
		\psi^{\alpha}_{\varepsilon\varkappa}(\mathbf{n})
	\right|^2
	\rho^\alpha ({\bf n}, \varepsilon).
\end{equation}
The latter equation is convenient for analysis of
$\Gamma_{\varepsilon\varkappa}$
in the limit of small energies
$|\varepsilon|\ll 3t$.
It is easy to prove that
$\rho^\alpha ({\bf n}, \varepsilon) \sim |\varepsilon|$ at small energy. This means that
$\Gamma_{\varepsilon\varkappa} \sim |\varepsilon|$.

To obtain a more qualitative estimate, we can use the fact that for small
$\varepsilon$ the localization length is large, and the square modulus of the edge state wave
function, $|\psi^{\alpha}_{\varepsilon\varkappa}(\mathbf{n})|^2$, decays slowly deep into the sample. Therefore, using the normalization condition $\sum_{\mathbf{n}\alpha}|\psi^{\alpha}_{\varepsilon\varkappa}(\mathbf{n})|^2=1$, we obtain approximately
\begin{eqnarray}
\sum_{\mathbf{n}\alpha}
	\left|
		\psi^{\alpha}_{\varepsilon\varkappa}(\mathbf{n})
	\right|^2
	\rho^\alpha ({\bf n}, \varepsilon)\approx\
	\frac{1}{N}\sum_{\mathbf{n}\alpha}\rho^\alpha ({\bf n}, \varepsilon)=\rho (\varepsilon)\,.
\end{eqnarray}
Combining this with
Eq.~\eqref{Gamma}
and with the formula~\cite{neto_etal}
$\rho(\varepsilon) \approx {|\varepsilon|}/{\sqrt{3}\pi t^2}$,
which is valid for small $\varepsilon$, we derive
\begin{eqnarray}
\Gamma_{\varepsilon\varkappa}
\approx
2\pi
n_{\text{imp}}V_0^2\rho(\varepsilon)
\approx
\frac{2 V_0^2 n_{\rm imp}}{\sqrt{3} t^2}|\varepsilon|\,.
\end{eqnarray}
Thus, for our edge states to be well-defined, we need:
\begin{eqnarray}
\Gamma_{\varepsilon\varkappa} \ll |\varepsilon|
\quad
\Leftrightarrow
\quad
V_0^2 n_{\rm imp} \ll t^2.
\end{eqnarray}
This condition serves as a definition of the weak disorder.

In addition to scattering into the bulk states, the edge states experience
the scattering into other edge states. In principle, such scattering leads
to the Anderson localization of the edge states. However, interplay of the
localization and the scattering into the bulk states has to be properly
investigated.

It is interesting to note that the
zigzag edge states are more resilient toward the disorder: they are located
at the zero energy, where the density of bulk states $\rho(\varepsilon)$
vanishes. In principle, the similar situation can take place in our case
too: for example, the functionalized edge with parameters $\varepsilon'=0$
and small $t_R\ll t$ guarantees that the edge band lies close to
$\varepsilon=0$, where the density of states in the bulk vanishes. The type
B solutions (for a wide range of model parameters) lie in the region of
energy where no bulk states exist. Thus, from Eq.~\eqref{Gamma} we obtain
$\Gamma_{\varepsilon\varkappa}=0$, and one can expect that these solutions
(for both types of edge modifications) are less sensitive to disorder. We
should mention, however, that Eq.~\eqref{Gamma} takes into account only
$p_z$ bulk electron band and neglect other graphene bands the edge
electrons can hybridize with. The detailed analysis of the effects of
disorder requires a separate study.

To conclude, we demonstrated that the armchair edge, when suitably altered,
supports edge states. We discussed two types of the edge modification: the
carbon-carbon hopping amplitude at the edge is unequal to the hopping
amplitude in the graphene bulk, and the chemically functionalized edge.
Both types stabilize the edge states, provided that the parameters are
suitably chosen. The properties of the edge state differ from the
properties of the edge states near zigzag termination, and depend on the
model parameters.

This work was supported by the Russian Foundation for Basic Research
(grants Nos.~11-02-00708, 11-02-91335, 11-02-00741, 12-02-92100, and
12-02-31400). A.O.S. and P.A.M. acknowledge support from the Dynasty Foundation.

\appendix

\begin{figure}
\includegraphics[width=\columnwidth]{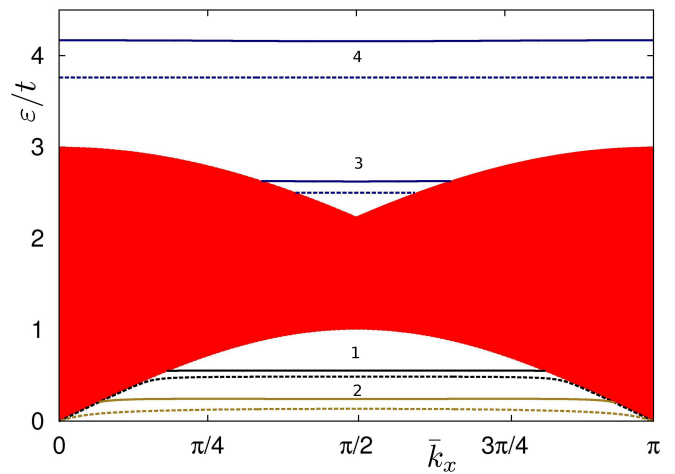}
 \caption{(Color online) Energy of localized electron states near the functionalized armchair edge calculated for four different set of parameters $\varepsilon'$ and $t_R$ in $t_R \ll t$ case. For each set of parameters, there are two edge bands $\varepsilon_{1,2}(k_x)$ (solid and dashed curves). The parameters are the following: (1) $\varepsilon'=0.5t$, $t_R=0.1t$ (type A), (2)  $\varepsilon'=0.2t$, $t_R=0.1t$ (type A), (3) $\varepsilon'=2.7t$, $t_R=0.1t$ (type B), (4) $\varepsilon'=3.7t$, $t_R=0.2t$ (type B). The (red) shaded area corresponds to the bulk graphene states with positive energy.}
 \label{p4energy}
\end{figure}

\section{Type A solution}\label{appendixA}
In the limit of small $t_R \ll t$ the solutions to Eq.~\eqref{p4eq1} exist only if $|\varepsilon'|\lesssim t$. There are two edge bands
\begin{equation}
\varepsilon_{1,2}(\bar{k}_x)=\varepsilon'+O(\tau_R^2)
\end{equation}
obeying the inequality:
\begin{equation}\label{varepsilon12}
\varepsilon_{1}(\bar{k}_x)<\varepsilon'<\varepsilon_{2}(\bar{k}_x)\,.
\end{equation}
The solutions can be written approximately as
\begin{equation}
\varepsilon_{1,2}(\bar{k}_x)=\varepsilon'+\tau_R^2\delta\varepsilon_{1,2}(\bar{k}_x),
\end{equation}
\begin{equation}
\bar{\varkappa} (\bar{k}_x) = {\varkappa}_0 (\bar{k}_x) + \tau_R^2 \delta \varkappa_{1,2} (\bar{k}_x)+ o(\tau_R^2),
\end{equation}
where ${\varkappa}_0 (\bar{k}_x)$ is given by the following equation:
\begin{equation}
\varepsilon(\bar{k}_x,{\varkappa}_0 (\bar{k}_x))=\varepsilon'\,.
\end{equation}
Using Eq.\eqref{ree2} we can write the expression for ${\varkappa}_0$:
\begin{equation}
\cosh \varkappa_0 (\bar{k}_x) =  \sqrt{\frac{(5-\frac{{\varepsilon'}^2}{t^2})+\sqrt{(5-\frac{{\varepsilon'}^2}{t^2})^2-16\cos^2 \bar{k}_x}}{8}}\,.
\label{kappa0}
\end{equation}
Expressions for $\delta \varkappa_{1,2}(\bar{k}_x)$ and ${\delta\varepsilon}_{1,2}(\bar{k}_x)$ are written as follows:
\begin{equation}
\delta\varepsilon_{1,2}(\bar{k}_x)=\frac{-\varepsilon'\mp\sqrt{4t^2\sinh^2 \varkappa_0+{\varepsilon'}^2 }}{e^{2\varkappa_0}-1},
\end{equation}
\begin{equation}
\delta \varkappa_{1,2} (\bar{k}_x)=-\frac{\varepsilon'\delta\varepsilon_{1,2}(\bar{k}_x)}{t^2\sinh(2\varkappa_0)\left[4-\displaystyle\frac{\cos^2\bar{k}_x}{\cosh^4\varkappa_0}\right]}\,.
\end{equation}
In contrast to the previous case ($t_R\gg t$) [see Fig.~\ref{p42en}] these solutions $\varepsilon_{1,2}(k_x)$ have weak dispersion. If $\varepsilon'\neq 0$ the particle-hole symmetry is violated: $\varepsilon_{2}(\bar{k}_x)\neq -\varepsilon_{1}(\bar{k}_x)$. Solutions of this type are shown in Fig.~\ref{p4energy} for two different sets of model parameters $t_R$($\ll t$) and $\varepsilon'$.

Finally, for $|\varepsilon'|\to\infty$ there is one solution to Eq.~\eqref{p4eq1} for any values of $t_R$.
\begin{equation}
\bar{\varkappa}\cong\frac{t_R^2|\sin\bar{k}_x|}{t|\varepsilon'|},
\end{equation}
\begin{equation}
 \varepsilon(\bar{k}_x)=\text{sgn}(\varepsilon')t|\sin(\bar{k}_x)|+O(\bar{\varkappa}^2).
 \end{equation}
As one can see from these formulas, similar to the case  $t_R \gg t$, this solution extends deeply into the bulk and has a pronounced dispersion.

\section{Type B solution}\label{appendixB}

 When $t_R \ll t$, as for the type A solution, there are two particle-hole non-symmetric bands, $\varepsilon_{1,2}(\bar{k}_x)$, satisfying the condition~\eqref{varepsilon12}. They also can be written in the form $\varepsilon_{1,2}(\bar{k}_x)=\varepsilon'+\tau_R^2\delta\varepsilon_{1,2}(\bar{k}_x)$, where $\delta\varepsilon_{1,2}(\bar{k}_x)$ are
 \begin{equation}
 \delta\varepsilon_{1,2}(\bar{k}_x)=t\frac{b\pm\sqrt{b^2-4a}}{2a},
 \end{equation}
 where
 
 \begin{align}
 a&=e^{ \bar{\varkappa}_{1}^{(0)}+ \bar{\varkappa}_{2}^{(0)}},\\
 b&=\frac{\varepsilon'(e^{ \bar{\varkappa}_{1}^{(0)}}+e^{ \bar{\varkappa}_{2}^{(0)}})}{\sqrt{\varepsilon'^2-t^2\sin^2 \bar{k}_x}}\,,
 \end{align}
and $\bar{\varkappa}_{1,2}^{(0)}(\bar{k}_x)$ are defined by
\begin{equation}
 \varepsilon(\bar{k}_x, \bar{\varkappa}_{1,2}^{(0)}(\bar{k}_x))=\varepsilon'\,,
\end{equation}

\begin{eqnarray}
\cosh\bar{\varkappa}_{1,2}^{(0)}(\bar{k}_x)&=&\mp  \frac{1}{2}\cos \bar{k}_x\\
&&+\frac{1}{2t}\sqrt{\varepsilon'^2-t^2\sin^2 \bar{k}_x}\,.\nonumber
\end{eqnarray}

 In contrast to the type A, the type B solutions exist if $|\varepsilon'|\gtrsim \sqrt{5}t$. These solutions have weak dispersion and lie above or below the bulk states. The results of numerical calculations for this case are shown in Fig.~\ref{p4energy}. Finally, when $|\varepsilon'| \gg t,t_R$, equation~\eqref{p4condition2} has two solutions with energies located near $\varepsilon'$. The edge state are strongly localized with $\bar{\varkappa}_{1,2}\approx\ln(\varepsilon'/t)$, and band spectra have a weak dispersion.

\bibliographystyle{apsrev_no_issn_url}

\bibliography{armchair_edge2}

\end{document}